\documentclass[10pt,aps.prl,twocolumn,superscriptaddress,floatfix]{revtex4-1} 

\usepackage{amsmath}
\usepackage{graphics}
\usepackage{color}
\usepackage{caption}
\usepackage{subcaption}
\usepackage{float}
\usepackage{array}
\usepackage{ulem}

\begin{document}

\title{Random Lasing and Reversible Photodegradation in Disperse Orange 11 Dye-Doped PMMA with Dispersed ZrO$_2$ Nanoparticles}
\author{Benjamin R. Anderson$^*$}
\author{Ray Gunawidjaja} 
\author{Hergen Eilers}

\affiliation{Applied Sciences Laboratory, Institute for Shock Physics, Washington State University,
Spokane, WA 99210-1495}
\date{\today}

\email{branderson@wsu.edu}

\begin{abstract}
We report the observation of intensity feedback random lasing at 645 nm in Disperse Orange 11 dye-doped PMMA (DO11/PMMA) with dispersed ZrO$_2$ nanoparticles (NPs). The lasing threshold is found to increase with concentration, with the lasing threshold for 0.1 wt\% being $75.8 \pm 9.4$ MW/cm$^2$ and the lasing threshold for 0.5 wt\% being $121.1 \pm 2.1$ MW/cm$^2$, with the linewidth for both concentrations found to be $\approx 10$ nm. We also consider the material's photostability and find that it displays fully reversible photodegradation with the photostability and recovery rate being greater than previously observed for DO11/PMMA without NPs. This enhancement in photostability and recovery rate is found to be explicable by the modified correlated chromophore domain model, with the NPs resulting in the domain free energy advantage increasing from 0.29 eV to 0.41 eV. Additionally, the molecular decay and recovery rates are found to be in agreement with previous measurements of DO11/PMMA [Polymer Chemistry \textbf{4}, 4948 (2013)]. These results present new avenues for the development of robust photodegradation-resistant organic dye-based optical devices.

\vspace{1em}
\end{abstract}



\maketitle

\vspace{1em}

\section{Introduction}
Lasing in scattering media -- known as random lasing (RL) -- was first predicted by Letokhov and coworkers \cite{Letokhov67.01,Letokhov67.02,Letokhov66.01} in the late 1960's and then experimentally observed by Lawandy \textit{et al.} in 1994 \cite{Lawandy94.01}.  RL differs from normal lasing in that random lasers operate without an external cavity, with scattering acting as the feedback mechanism \cite{Cao03.01,Wiersma96.01,Wiersma08.01}.  Due to different scattering regimes in diffuse media, RL is found to have two distinct spectral classes: intensity feedback random lasing (IFRL) and resonant feedback random lasing (RFRL) \cite{Cao03.01,Cao05.01,Ignesti13.01}.  IFRL is characterized by a single narrow emission peak (FWHM on the order of 10's of nm) and is wholly determined by the diffusive nature of light \cite{Lawandy94.01,Burin01.01,Wiersma96.01,Pinheiro06.01}.  On the other hand, RFRL is characterized by multiple sub-nm width peaks \cite{Ling01.01,Cao03.01,Cao03.02,Cao05.02,Cao99.01,Tureci08.01} with 
two proposed mechanisms:  strong scattering resonances \cite{Molen07.01} and Anderson localization of light \cite{Cao00.01}.  Based on these proposed mechanisms, models of RFRL have been developed using spin-glass modeling of light \cite{Angelani06.01},  Levy-flight scattering \cite{Ignesti13.01}, condensation of lasing modes \cite{Conti08.01,Leonetti13.03}, and strongly interacting lossy modes \cite{Tureci08.01}.

Regardless of the microscopic  mechanisms of the two regimes, their spectral characteristics can be described macroscopically in terms of active lasing modes, with RFRL representing a few distinct active lasing modes, and IFRL representing multiple overlapping active lasing modes \cite{Cao03.01,Cao03.02,Ling01.01,Tureci08.01,Ignesti13.01}.  The different modal nature of the two regimes makes each attractive for different applications.  Since RFRL has few modes --allowing for the creation of unique spectral signatures -- it is attractive in the fields of authentication \cite{Zurich08.01}, biological imaging, emergency beacons \cite{Hoang10.01,Cao05.01}, and random number generation \cite{Atsushi08.01,Murphy08.01,Mgrdichian08.01}.  Also, the limited number of active modes in RFRL allows for a high degree of spectral control, as the pump beam can be modulated using a spatial light modulator (SLM) to activate only certain lasing modes, thus controlling the RFRL spectrum \cite{Leonetti12.01,Cao05.01,Leonetti13.02,Leonetti12.02,Leonetti12.03,Andreasen14.01,Bachelard12.01,Bachelard14.01}.  This spectral control is attractive for implementing optically based physically unclonable functions  \cite{Anderson14.04,Anderson14.05,Eilers14.01} and the creation of bright tunable light sources \cite{Cao05.01}.

In the case of IFRL, the many active modes leads to the emission having a low degree of spatial coherence \cite{Redding11.01}, making it an attractive method for high-intensity low-coherence light sources \cite{Redding12.01}.  Such light sources have applications in photodynamic therapy, tumor detection \cite{Hoang10.01,Cao05.01}, flexible displays, active elements in photonics devices \cite{Cao05.01}, picoprojectors, cinema projectors \cite{Hecht12.01}, and biological imaging \cite{Redding12.01,Hecht12.01}.


With the wide variety of possible applications for RL, work on RL has focused on discovering materials that have the following three characteristics: they provide desirable RL spectra, are relatively cheap and easy to work with, and are robust enough to use over a reasonable time frame. One such class of materials that fulfills the first two criteria are organic-dye-based materials. Unfortunately, most organic-dye-based systems are found to \textit{irreversibly} photodegrade when exposed to intense radiation \cite{wood03.01,taylo05.01,Avnir84.01,Knobbe90.01,Kaminow72.01,Rabek89.01}, thus limiting their usefulness in optical devices.  However, in the past two decades it has been discovered that some dye-doped polymers actually photodegrade \textit{reversibly}, with the material self-healing once the illumination is turned off for a period of time. These materials include Rhodamine B and Pyrromethene dye-doped (poly)methyl-methacrylate (PMMA) optical fibers \cite{Peng98.01}, disperse orange 11 (DO11) dye-doped 
PMMA \cite{howel04.01,howel02.01} and styrene-MMA copolymer \cite{Hung12.01}, anthraquinone-derivative-doped PMMA \cite{Anderson11.02}, 8-hydroxyquinoline (Alq) dye-doped PMMA \cite{Kobrin04.01}, and air force 455 (AF455) dye-doped PMMA \cite{Zhu07.01}.  In all these studies the dye was doped into the polymer without any scattering particles, such that no RL was observed.

Given the large number of dye-doped polymers that display self-healing (but not RL) and the desirability of a self-healing organic dye-based random laser, we recently tested random lasers consisting of Rhodamine 6G dye-doped polyurethane with dispersed ZrO$_2$ (R6G+ZrO$_2$/PU) \cite{Anderson14.04} or Y$_2$O$_3$ (R6G+Y$_2$O$_3$/PU) nanoparticles (NP) for reversible photodegradation \cite{Anderson15.01,Anderson15.03}. In those studies we found that R6G+ZrO$_2$/PU and R6G+Y$_2$O$_3$/PU display self-healing after photodegradation, with a recovery efficiency of 100\% \cite{Anderson15.01,Anderson15.03}. However, we also found that the photodegradation could not be called truly reversible as the RL wavelength and linewidth changed \cite{Anderson15.01,Anderson15.03}.  

While the approach used in our recent study -- of testing an already known RL system for self-healing -- was successful at producing a self-healing organic dye based random laser, a different approach to the problem is to develop an already known self-healing system into a random laser. To this end we investigate RL in DO11/PMMA with dispersed ZrO$_2$ NPs.  The choice to use DO11/PMMA is three-fold: (1) DO11 has previously been shown to be suitable as a laser dye \cite{howel02.01,howel04.01}, (2) the majority of organic-dye based RL studies have focused on Rhodamine dyes and therefore DO11 is a new and unique organic dye in RL studies, with its lasing wavelength being attractive for use with polymer optical fibers \cite{howel02.01} and (3) DO11/PMMA is the test bed system for self-healing research, with numerous studies performed to understand the phenomenon of self healing in DO11/PMMA.

These studies have been performed with different probe techniques including: absorption  \cite{embaye08.01,Anderson14.02}, white light interferometry \cite{Anderson14.03},  fluorescence \cite{Dhakal12.01,raminithesis}, photoconductivity \cite{Anderson13.02}, transmittance microscopy \cite{Anderson11.01, Anderson13.01}, and amplified spontaneous emission (ASE) \cite{howel02.01,howel04.01,embaye08.01,Ramini12.01,Ramini13.01}.  These techniques have been used to characterize the behavior of DO11/PMMA's photodegradation and recovery under different wavelengths \cite{Anderson15.04}, temperatures \cite{Ramini12.01,Ramini13.01,raminithesis,andersonthesis}, applied electric fields \cite{Anderson13.01,andersonthesis,Anderson14.01}, co-polymer compositions \cite{Hung12.01}, thicknesses \cite{Anderson14.01}, concentrations  \cite{Ramini12.01,Ramini13.01,raminithesis}, and intensities \cite{Anderson11.01,Anderson14.01,Anderson14.02}. Based on all these studies a model has been developed to describe DO11/PMMA's photodegradation and recovery called the correlated chromophore domain model (CCDM) \cite{Ramini12.01,Ramini13.01,raminithesis,Anderson14.02,andersonthesis}. 

The CCDM posits that dye molecules form linear isodesmic domains along polymer chains with molecular interactions -- mediated by the polymer -- resulting in increased photostability and self-healing.  Within the domain model the decay rate $\alpha$, depends inversely on the domain size $N$, as

\begin{equation}
\alpha=\frac{\alpha_1}{N},\label{eqn:domdec}
\end{equation}
and the recovery rate $\beta$, depends linearly on the domain size,
\begin{equation}
\beta=\beta_1N,\label{eqn:domrec}
\end{equation}
where $\alpha_1$ and $\beta_1$ are the unitary domain decay and recovery rates, respectively.  While these rates describe the dynamics of a single domain, the macroscopically measured rates result from an ensemble average over the distribution of domains  $\Omega(N)$, which depends on the density of dye molecules $\rho$, and the free energy advantage $\lambda$ \cite{Ramini12.01,Ramini13.01,raminithesis,Anderson14.02,andersonthesis}.

\section{Method}
In order to produce a suitable DO11 based random lasing material we disperse ZrO$_2$ NPs into DO11 dye-doped PMMA.  We begin by first fabricating the ZrO$_2$ NPs using forced hydrolysis followed by calcination at a temperature of 600 $^{\circ}$C for an hour \cite{Gunawidjaja13.01}. The ZrO$_2$ NPs are then functionalized by dispersing them in a 2.5 vol\% solution of 3-(Trimethoxysilyl)propyl methacrylate in toluene, which is subsequently refluxed for 2 h \cite{Gunawidjaja11.01}. To prepare the dye-doped polymer, we first filter Methyl methacrylate (MMA)  through a column of activated basic alumina to remove inhibitor.  Next we dissolve 25 wt\% PMMA into the inhibitor-free MMA and divide the MMA/PMMA solution into three batches for different dye concentrations.  DO11 dye (TCI America, purity $>$98\%) is added to the MMA/PMMA solution in concentrations of 0.1 wt\%, 0.5 wt\%, and 1.0 wt\%.  The functionalized ZrO$_2$ NPs are then added at a concentration of 10 wt\% and the mixture is sonicated until it is homogeneous, at which point 0.25 wt\% 2, 2'-azobis(2-methyl-propionitrile) is added and the mixture is further sonicated before being poured onto 1''$\times$1.5'' glass slides.  The samples are then covered and placed in an oven at 60-65 $^\circ$C for 2 h to cure. Once prepared the samples are characterized using SEM, absorption spectroscopy, transmission measurements, and mechanical measurements.  The results of the relevant sample parameters are tabulated in Table \ref{tab:param}.

\begin{figure}
 \centering
 \includegraphics{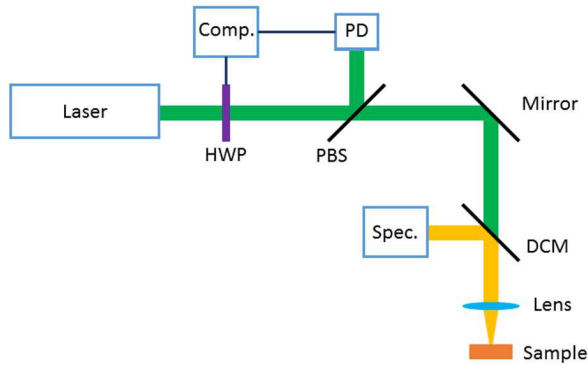}
 \caption{Schematic of RL setup.}
 \label{fig:setup}
\end{figure}

To measure the sample's emission we use an intensity controlled random lasing system \cite{Anderson15.03} shown schematically in Figure \ref{fig:setup}.  The pump is a Spectra-Physics Quanta Ray Pro Q-switched frequency doubled Nd:YAG laser (532 nm, 10 Hz, 10 ns) with the emission stabilized using a motorized half-waveplate (HWP) and polarizing beamsplitter (PBS) combination with a Thorlabs Si photodiode (PD) providing feedback for the HWP.  The stabilized pump beam is focused onto the sample using a spherical lens with a focal length of 50 mm.  Once pumped, the sample emits light in the backward direction, which is collimated using the focusing lens and then reflected by a dichroic mirror (DCM) (cutoff wavelength of 550 nm) into an optical fiber connected to a Princeton Instruments Acton 2300i spectrometer with a Pixis 2K CCD detector. For reference the relevant experimental parameters are tabulated in Table \ref{tab:param}.

\begin{table}
\centering
\caption{Relevant sample and experimental parameters as proposed by van der Molen \textit{et al.} \cite{Molen07.02}. $l$: transport mean-free path, $l_a$: ballistic absorption length, $d_{NP}$: average NP diameter, $\rho_{NP}$: density of NPs, $L$: nominal sample thickness, $\lambda_p$: pump wavelength, $r_p$: pump repetition rate, $\Delta t$: pulse width, $A$: Spot size, $\Delta \lambda$: spectral resolution.}
\label{tab:param}
\begin{tabular}{|ccl|}
\hline
\multicolumn{3}{|c|}{\textbf{Sample Parameters}} \\ \hline
$l$$^1$ & $4.10 \pm 0.25$ & $\mu$m \\
$l_a$$^2$ & $657\pm 20$ & $\mu$m \\
$d_{NP}$$^3$ &$195 \pm 32$ &nm \\
$\rho_{NP}$& $2.26 \times 10^{12}$ & cm$^{-3}$\\ 
$L$ & $\approx 500$  & $\mu$m \\ \hline
\multicolumn{3}{|c|}{\textbf{Experimental Parameters}} \\ \hline
$\lambda_p$ &532 & nm \\ 
$r_p$ & 10 & Hz \\
$\Delta t$ & 10 & ns \\
$A$ & $7.85\times10^{-3}$ & cm$^2$\\
$\Delta\lambda$ & 0.27 & nm\\ \hline
\multicolumn{3}{l}{$^1$Calculated using the total integrated } \\
\multicolumn{3}{l}{transmission method \cite{Rivas99.01,Vellekoop10.02}} \\
\multicolumn{3}{l}{$^2$For a dye concentration of 0.1 wt\%} \\
\multicolumn{3}{l}{$^3$The NP size and distribution was}\\
\multicolumn{3}{l}{previously characterized using}\\
\multicolumn{3}{l}{an SEM in Ref. \cite{Anderson15.02}.}\\
\end{tabular}
\end{table}

\section{Results and discussion}
\subsection{Random Lasing Properties}
\begin{figure}
 \centering
 \includegraphics{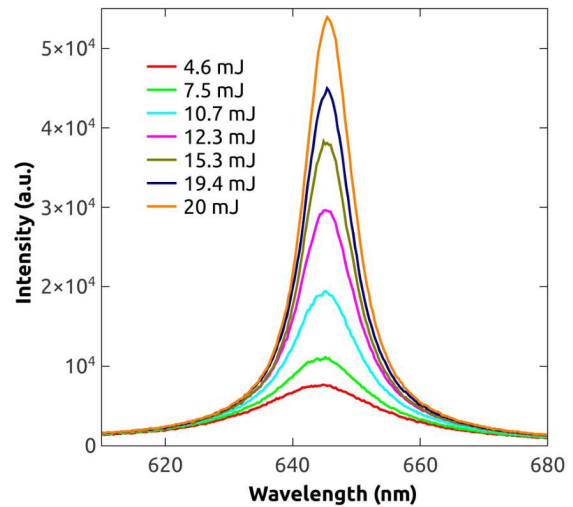}
 \caption{Random lasing spectra as a function of wavelength for different pump energies for a dye concentration of 0.1 wt\%.}
 \label{fig:RLspec}
\end{figure}

\begin{figure}
 \centering
 \includegraphics{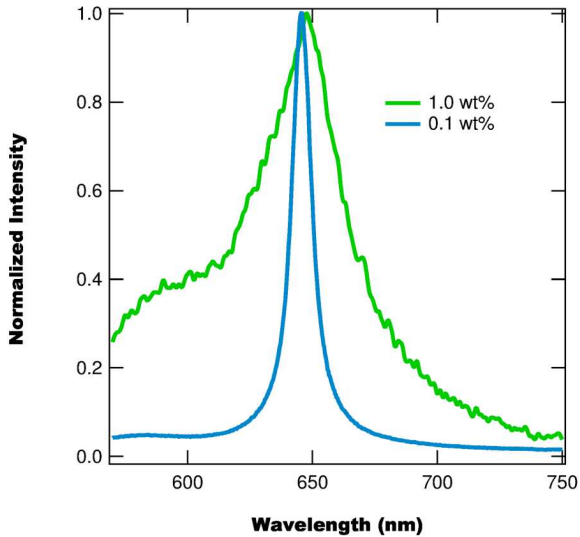}
 \caption{Comparison of emission from 0.1 wt\% sample and 1.0 wt\% sample. Note that the 0.1 wt\% sample is pumped with an energy of 10 mJ, while the 1.01 wt\% sample is pumped with an energy of 60 mJ, which is near the ablation threshold.}
 \label{fig:denscomp}
\end{figure}

We test DO11+ZrO$_2$ for RL using a NP concentration of 10 wt\% and three different dye concentrations (0.1 wt\%, 0.5 wt\%, and 1.0 wt\%) and find that the 0.1 wt\% and 0.5 wt\% samples display RL, while the 1.0 wt\% samples are not found to produce RL.  Figure \ref{fig:RLspec} shows the emission spectra for the 0.1 wt\% sample at several pump energies, with the emission narrowing into a single lasing peak as the pump energy passes the lasing threshold, while Figure \ref{fig:denscomp} compares the normalized spectra of the 0.1 wt\% sample (pump energy of 10 mJ) and the 1.0 wt\% sample (pump energy of 60 mJ). Note that the 1.0 wt\% sample is pumped at a level near it's ablation threshold (i.e. any higher pump energies result in the material being ablated by a single pulse). From Figure \ref{fig:denscomp} we find that at high pump energies the emission from the 0.1 wt\% sample is characteristic of IFRL, while the emission from the 1.0 wt\% sample is much broader. From the spectral shape of the 1.0 wt\% emission, and it's peak location of 647.5 nm, we conclude that the emission corresponds to a combination of amplified spontaneous emission (ASE) and fluorescence, as DO11/PMMA's ASE wavelength is known to be $\approx 650$ nm \cite{howel02.01,howel04.01,embaye08.01}.  Since we are primarily concerned with RL in DO11+ZrO$_2$/PMMA, the remainder of this study will focus only on the samples with dye concentrations of 0.1 wt\% or 0.5 wt\%.

\begin{figure}
  \centering
  \includegraphics{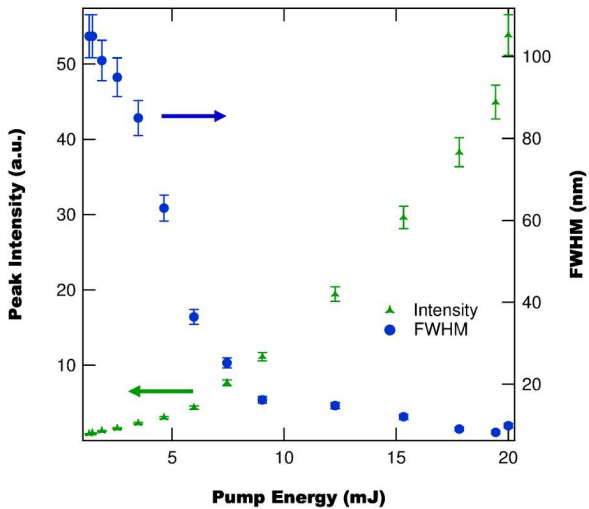}
  \caption{Peak intensity and linewidth as a function of pump energy for a sample with a dye concentration of 0.1 wt\% and NP concentration of 10 wt\%.  From both the peak intensity and linewidth we determine a lasing threshold of $75.8 \pm 9.4$ MW/cm$^2$.}
  \label{fig:01}
 \end{figure}

\begin{figure}
  \centering
  \includegraphics{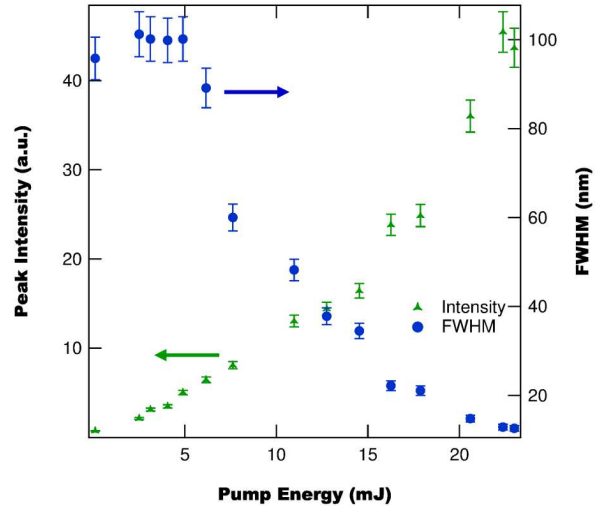}
  \caption{Peak intensity and linewidth as a function of pump energy for a sample with a dye concentration of 0.5 wt\% and NP concentration of 10 wt\%.  From both the peak intensity and linewidth we determine a lasing threshold of $121.1 \pm 2.1$ MW/cm$^2$.}
  \label{fig:05}
 \end{figure}

We characterize the RL properties of the 0.1 wt\% and 0.5 wt\% samples by considering three RL features: peak intensity, peak wavelength, and RL linewidth (e.g. FWHM).  Figure \ref{fig:01} shows the peak intensity and linewidth for the 0.1 wt\% sample, while Figure \ref{fig:05} shows the same quantities for the 0.5 wt\% sample.  From Figure \ref{fig:01} we see that the transition to RL for the 0.1 wt\% sample is quick, with the linewidth narrowing from 100 nm at 3 mJ to 10 nm at 9 mJ, and the intensity's slope changes by a factor of $\approx 4.1\times$ once above the lasing threshold. While the transition for the 0.1 wt\% sample is found to be abrupt, the transition to RL for the 0.5 wt\% sample is more gradual.  From Figure \ref{fig:05} we find that the linewidth changes from 100 nm at 5 mJ to 10 nm at 25 mJ and the intensity's slope only increases by a factor of $\approx 2.7\times$. This more gradual transition into lasing suggests that there is more competition between ASE and lasing \cite{Andreasen10.01,Cao00.02} in the 0.5 wt\% sample, than 
in the 0.1 wt\%.


While Figures \ref{fig:01} and \ref{fig:05} can help us understand the underlying spectral properties of the sample's emission, they also can be used to directly determine the sample's lasing threshold.  Using either the FWHM as a function of pump energy \cite{Cao03.01} or a bilinear fit to the peak intensity \cite{Vutha06.01, Anderson14.04} the lasing threshold of each sample can be calculated with the 0.1 wt\% sample having a lasing threshold of $75.8 \pm 9.4$ MW/cm$^2$ and the 0.5 wt\% sample having a threshold of $121.1 \pm 2.1$ MW/cm$^2$.  Note that these thresholds are much larger ($\approx 10\times$) than similar RL materials based on R6G \cite{Anderson14.04}. These large lasing thresholds are due to DO11 having a smaller gain coefficient than R6G \cite{howel02.01,howel04.01,Mysliwiec09.01} as well as our use of off resonance pumping ($\lambda_{pump}=532$ nm and $\lambda_{res}=470$ nm).

Based on the observed lasing thresholds -- and the observation that the 1.0 wt\% sample didn't lase even with a pump energy of 60 mJ ($I=754$ MW/cm$^2$) -- we find that the RL threshold of DO11+ZrO$_2$/PMMA increases with dye concentration, which is counter to measurements in other dyes \cite{Anderson14.04}.  One possible explanation for this effect is the formation of dimers at higher concentrations.  From studies in other organic dye materials it is known that dimer formation leads to a redshift in the absorption spectrum of the material \cite{Toudert13.01,Gavrilenko06.01,Arbeloa88.01}.  This redshift can result in fluorescence quenching \cite{Penzkofer86.01,Penzkofer87.01,Bojarski96.01,Setiawan10.01}, which will decrease the material's RL gain leading to the RL threshold increasing.  Additionally, dimer formation can describe the difference in the shape of the linewidth as a function of pulse energy (i.e. the 0.1 wt\% has a sudden decrease in linewidth and the 0.5 wt\% has a slow decrease) as a similar effect has been observed in comparisons of RL in Rhodamine B monomers and dimers \cite{Marinho15.01}.

The last RL property we consider is the peak wavelength as a function of pump energy, which is shown in Figure \ref{fig:wave} for both the 0.1 wt\% and the 0.5 wt\% samples.  Both sample's begin with their peak emission near 625 nm, with the peak wavelength smoothly transitioning into a steady lasing wavelength after the lasing threshold. The 0.1 wt\% sample is found to have its RL peak at 645 nm, while the 0.5 wt\% sample is found to have its RL peak at 646 nm.  These two results, along with the observation of the 1.0 wt\% sample's ASE wavelength of 647.5 nm suggests that as the dye concentration increases the peak emission is redshifted.  This is a known effect caused by increased self-absorption due to the greater dye concentration \cite{Shuzhen09.01,Ahmed94.01,Shank75.01} and subsequent dimer formation \cite{Toudert13.01,Gavrilenko06.01,Arbeloa88.01}.

\begin{figure}
 \centering
 \includegraphics{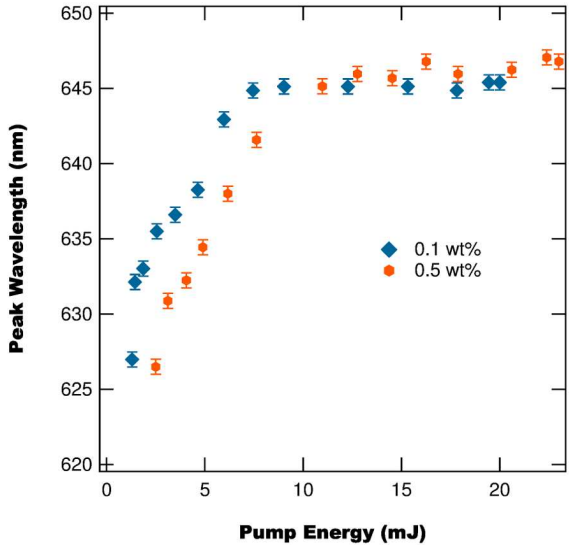}
 \caption{Peak wavelength as a function of pump energy for the 0.1 wt\% and 0.5 wt\% dye concentration samples. The low intensity emission is peaked near 625 nm and smoothly transitions with increasing pump energy to be centered at $\approx 645$ nm.}
 \label{fig:wave}
\end{figure}

\subsection{Photodegradation and self-healing}
\begin{figure}
 \centering
 \includegraphics{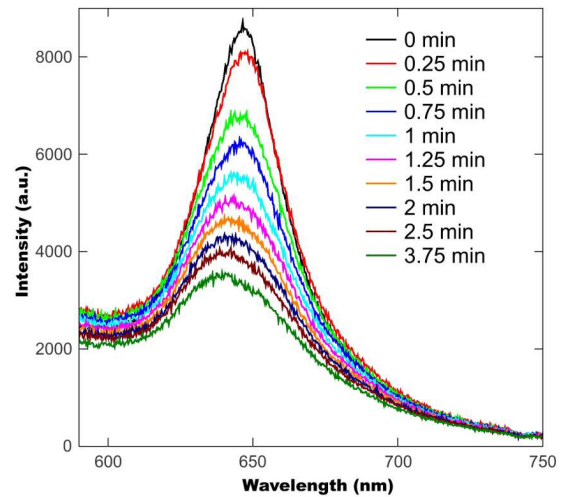}
 \caption{Random lasing spectra as a function of wavelength at different times during photodegradation for a pump energy of 7 mJ and a 0.1 wt\% dye-concentration sample.}
 \label{fig:dec}
\end{figure}

With DO11+ZrO$_2$/PMMA found to lase in the IFRL regime for low dye concentrations, we now turn to consider the effect of ZrO$_2$ NPs on DO11/PMMA's ability to self heal.  For these measurements we use a sample with a dye concentration of 0.1 wt\% and a NP concentration of 10 wt\%. We use a 7 mJ/pulse (time averaged intensity of $I_{avg}=8.9$ W/cm$^2$) beam for both degrading the sample and measuring the RL spectra.  During decay, the beam is always incident on the sample, while during recovery the beam is blocked except when taking measurements of the sample's RL spectrum. Spectral measurements during recovery involve exposing the sample to three pump pulses to determine the average RL spectrum. These measurements occur every ten minutes during recovery, which equates to a duty cycle of 0.05\%.  Figure \ref{fig:dec} shows the measured RL spectra during decay at several times, with the peak blueshifting and becoming broader.  The large background fluorescence in Figure \ref{fig:dec} is due to pumping the sample only slightly above it's lasing threshold (7 mJ pump, 5.9 mJ threshold).

\begin{figure}
 \centering
 \includegraphics{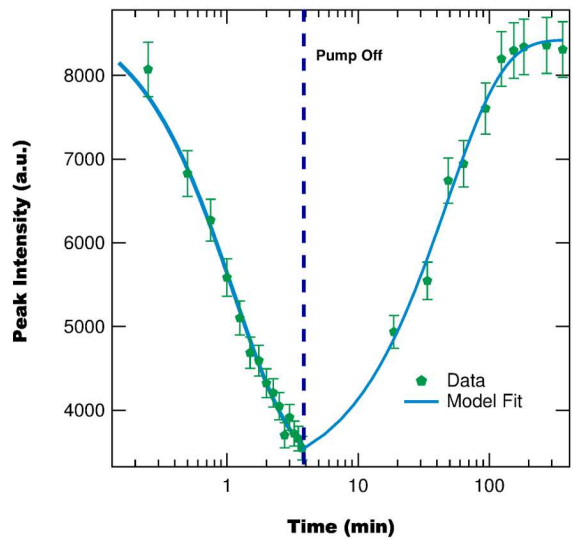}
 \caption{Peak intensity as a function of time during decay and recovery.}
 \label{fig:pint}
\end{figure}

From the spectra recorded during decay and recovery we determine the peak emission intensity as a function of time, shown in Figure \ref{fig:pint}.  The peak intensity is found to decay to 40\% of its initial value during degradation and found to fully recover (within uncertainty) after the pump beam is turned off.  This observation is consistent with ASE measurements of DO11/PMMA without dispersed NPs \cite{howel04.01,howel02.01,embaye08.01}, where the ASE signal is found to fully recover after degradation.

While we observe full reversibility for a degree of decay of up to 60\%, we also perform decay measurements with extreme degrees of decay ($\approx 95$ \%) and observe partial recovery. This suggests that as the degree of degradation increases past some threshold value full recovery is lost and the material incurs some irreversible damage.  To quantify this threshold degree of degradation at which point reversibility is lost we are planning measurements to systematically vary the degree of degradation and measure the degree of recovery.

In addition to performing preliminary measurements of how reversible photodegradation changes with the degree of damage, we also consider how cycling through degradation and recovery effects the material's self-healing.  These measurements so far have consisted of two decay and recovery cycles with full reversibility observed in both cycles, which is consistent with DO11/PMMA without disperesed NPs \cite{howel02.01}. Further work is planned to consider how many cycles can be completed before full reversibility is lost.

\subsubsection{Random Lasing Intensity Decay and Recovery}

To further characterize the influence of the dispersed NPs on DO11/PMMA's photodegradation and self-healing we determine the decay and recovery parameters by fitting the peak intensity as a function of time to a simple model of the RL intensity.  Assuming that only undamaged molecules (with fractional number density $n(t)$) participate in RL, the material's laser gain will be proportional to $n(t)$ leading to a RL intensity of \cite{Menzel07.01},  

\begin{equation}
  I_{RL}(t)=I_0e^{\sigma n(t)},\label{eqn:int}
\end{equation}
where $I_0$ is the initial peak intensity and $\sigma$ is the RL cross section. To model the population dynamics of the molecules we use Embaye \textit{et al.}'s two-species non-interacting molecule model, in which undamaged molecules reversibly transition into a damaged state during degradation \cite{embaye08.01}.  In this model the undamaged population's fractional number density during decay ($t\leq t_D$) is

\begin{equation}
 n(t)=\frac{\beta}{\beta+\alpha I}+\frac{\alpha I}{\beta+\alpha I}e^{-(\beta+\alpha I)t},
\end{equation}
 and during recovery ($t>t_D$) is,
 
\begin{equation}
 n(t)=1-[1-n(t_D)]e^{-\beta(t-t_D)},\label{eqn:rec}
\end{equation}
where $t_D$ is the time at which the pump is turned off, $\alpha$ is the decay parameter, $I$ is the pump intensity, and $\beta$ is the recovery rate.  Using Equations \ref{eqn:int}--\ref{eqn:rec} we can model the RL peak intensity's decay and recovery as a function of time and extract the relevant dynamical parameters, with  $\alpha=3.16(\pm 0.10) \times 10^{-2}$ cm$^2$W$^{-1}$min$^{-1}$ and $\beta =3.75( \pm 0.18) \times 10^{-2}$ min$^{-1}$.  The decay parameter, $\alpha$, is found to be smaller than the previously measured values for DO11/PMMA \cite{howel02.01,howel04.01,embaye08.01,Ramini12.01,Ramini13.01}, which means that the addition of nanoparticles improves the materials photostability. Additionally, we find that the recovery rate of DO11+ZrO$_2$/PMMA is larger than any previous measurement \cite{howel02.01,howel04.01,embaye08.01,Anderson11.01,Anderson13.01,Anderson14.01,Anderson14.02,Ramini12.01,Ramini13.01,Anderson15.04}, implying that the addition of ZrO$_2$ NPs helps aid the recovery process.  An explanation for these effects is that the introduction of NPs can change the free energy advantage $\lambda$, and density parameter, $\rho$, such that the average domain size is greater with NPs than without \cite{Ramini12.01, Ramini13.01,Anderson14.02}.    

While a precise determination of the domain parameters is beyond the scope of the current study, we can estimate the modified domain parameters by considering the effect of the average domain size on the recovery rates.  Previously it was shown that the average domain size is given by \cite{Anderson14.02}:

\begin{equation}
\langle N \rangle=\frac{\beta_M\Omega_1(\rho,\lambda)(1+z\Omega_1(\rho,\lambda))}{\rho |z\Omega_1(\rho,\lambda)-1|^3}, \label{eqn:N}
\end{equation}
where $\Omega_1(\rho,\lambda)$ is the density of unitary domains and $z=\exp\{\lambda/kT\}$ with $T$ being the temperature and $k$ being Boltzmann's constant. Using the average domain size (Equation \ref{eqn:N}) and Equations \ref{eqn:domdec}  and \ref{eqn:domrec} the measured decay and recovery rates can be approximated as

\begin{eqnarray}
\alpha&\approx\frac{\alpha_1}{\langle N\rangle}, \label{eqn:alp}
\\\beta&\approx \beta_1\langle N \rangle,\label{eqn:bet}
\end{eqnarray}
where once again $\alpha_1$ and $\beta_1$ are the unitary domain decay and recovery rates, respectively.

Therefore, assuming the unitary domain recovery rate is the same for DO11/PMMA both with and without NPs, we can determine the ratio of average domain sizes by taking the ratio of recovery rates between a sample with NPs and a sample without NPs:

\begin{eqnarray}
\frac{\beta}{\beta_0}=\frac{\langle N\rangle}{\langle N_{0}\rangle}
\\ =\frac{\rho_0\Omega_1(\rho,\lambda)}{\rho\Omega_1(\rho_0,\lambda_0)}\frac{1+z\Omega_{1}(\rho,\lambda)}{1+z_0\Omega_1(\rho_0,\lambda_0)}\frac{ |z_0\Omega_1(\rho_0,\lambda_0)-1|^3}{ |z\Omega_1(\rho,\lambda)-1|^3}\label{eqn:ratio}
\end{eqnarray}
where the subscript 0 corresponds to the parameters for the system with no nanoparticles and no subscript corresponds to the system with nanoparticles.  Comparing the recovery rate of DO11+ZrO$_2$/PMMA to a similarly dye-doped DO11/PMMA without NPs  \cite{Ramini13.01,raminithesis} we find a ratio of $\beta/\beta_0\approx 12.5$, which means that with the inclusion of NPs the domain size is an order of magnitude larger. Given this large difference, and the linear relationship between domain size and the density parameter, we conclude that primary influence of the NPs is on the free energy advantage.  Assuming that the density parameter is unchanged by the introductions of NPs, we can numerically solve Equation \ref{eqn:ratio} for the new free energy advantage and find $\lambda \approx 0.41$ eV, which is 0.12 eV larger than DO11/PMMA's value of 0.29 eV \cite{Ramini12.01,Ramini13.01,Anderson14.02}.

With the new free energy advantage determined, we can estimate the average domain size for our system and the unitary domain decay and recovery rates.  Substituting the new free energy advantage into Equation \ref{eqn:N} we find that the average domain size of our system is $\langle N \rangle = 375$. Using this domain size, along with Equations \ref{eqn:alp} and \ref{eqn:bet}, we determine the unitary domain decay rate to be $\alpha_1\approx 11.86 \pm 0.38 $ cm$^2$W$^{-1}$min$^{-1}$ and the unitary domain recovery rate to be $\beta_1\approx 1.00(\pm0.10)\times 10^{-4}$ min$^{-1}$, which are both found to be within uncertainty of the measured values for DO11/PMMA \cite{Ramini13.01}. This agreement of the unitary domain decay and recovery rate between DO11/PMMA and DO11+ZrO$_2$/PMMA implies that we are correct in our original assumption that the NPs only influence the domain size (via the free energy advantage) and do not influence the underlying molecular interactions leading to reversible photodegradation. Additionally, the success of the CCDM to correctly account for the influence of NPs on DO11/PMMA's decay and recovery is a strong indication that the CCDM is a robust description of reversible photodegradation for DO11 dye-doped polymers, with the addition of NPs resulting in the free energy advantage increasing.

One possible explanation for this increase in free energy advantage is that the introduction of NP's affects the local electric field experienced by the dye molecules, thereby influencing the underlying interactions behind the free energy advantage.  This effect on the local electric field arises due to the introduction ZrO$_2$ NP's (with a dielectric constant of $\approx 4.88$) into the polymer (with dielectric constant $\approx 2.22$) resulting in an increased dielectric constant of the dye's local environment. To estimate the magnitude of this effect on the local field factor we recall that for a spherical cavity in a uniform dielectric medium, with dielectric constant $\epsilon$, the local field factor is given by \cite{jacks96.01}:

\begin{equation}
L\propto \frac{3\epsilon}{2\epsilon+1}.\label{eqn:LFF}
\end{equation}
Using the relevant concentrations of NPs, dye, and polymer we determine that the permitivity is approximately 10\% larger, which using Equation \ref{eqn:LFF} results in the local field factor becoming 3\% larger. Since the dielectric energy of a system depends on the square of the local field factor, we conclude that the dielectric influence of the NPs increases the free energy advantage by 6\%, or 0.018 eV, which is too small to account for the total change in energy calculated above. This suggests two different possibilities: (1) the change in the free energy advantage also includes a contribution from another unidentified effect, most likely related to the interactions of dye, polymer, and NPs or (2) our simplistic treatment of the dielectric constant and local field factor may underestimate the actual enhancement in the local electric field, especially since ZrO$_2$ is a transparent conductive oxide \cite{Brune98.01,Naik13.01}, which can lead to plasmonic effects that can drastically increase the electric field \cite{Naik13.01}.

\subsubsection{Effect of photodegradation and recovery on linewidth and wavelength of random lasing} 

In addition to considering the decay and recovery dynamics of the RL intensity we also quantify the changes in the lasing peak location and linewidth during decay and recovery, as shown in Figure \ref{fig:wave}.  From Figure \ref{fig:wave} we find that during degradation the lasing peak blueshifts and the FWHM increases. After the pump beam is blocked (except for measurements during recovery), both the lasing peak and FWHM are found to return to within uncertainty of their initial values, suggesting that the photodegradation process is truly \textit{reversible}.  This result is different than observed for R6G+NP/PU, where the peak intensity fully recovers, but the lasing wavelength and linewidth are irreversibly changed due to photodegradation \cite{Anderson15.01,Anderson15.03}.


\begin{figure}
 \centering
 \includegraphics{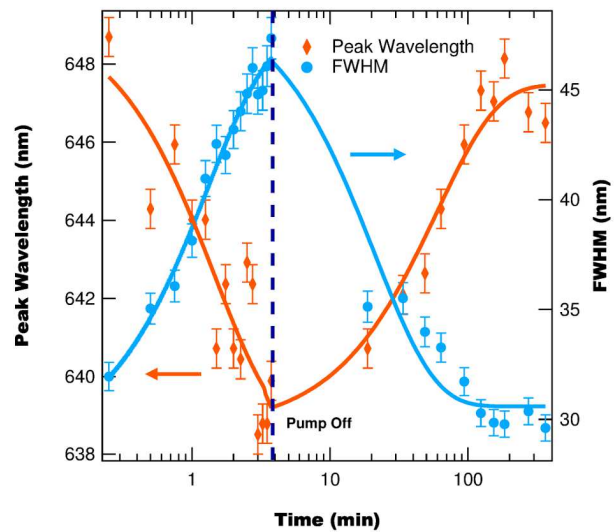}
 \caption{Peak wavelength and RL linewidth as a function of time during decay and recovery.}
 \label{fig:wave}
\end{figure}

The other difference, observed in Figure \ref{fig:wave}, between the photodegradation and self-healing of DO11+ZrO$_2$/PMMA and R6G+NP/PU, is that the lasing wavelength is found to blueshift during decay for DO11+ZrO$_2$/PMMA, while it is found to redshift for R6G+NP/PU.  In R6G+NP/PU, the redshifting of the lasing peak during photodegradation and recovery is attributed to photothermal heating and the formation of R6G dimers and trimers \cite{Anderson15.01,Anderson15.03}.  The observation of the opposite effect in DO11+ZrO$_2$/PMMA suggests a different mechanism responsible, with the most likely mechanism being related to the observation of a blueshift in DO11/PMMA's absorbance spectrum during photodegradation \cite{embaye08.01,Anderson13.01,andersonthesis,Anderson14.02}.  The effect of a blueshift in the absorbance spectrum leads to the shorter wavelengths experiencing less loss which results in RL emission blueshifitng, such that the gain-to-loss ratio is maximized.  This effect is also observed in the ASE spectrum of DO11 in different solvents, where a blueshift in the absorbance peak leads to a blueshift in the emission peak \cite{howel04.01}.  Additionally, this blueshift will lead to a larger portion of the emission spectrum being amplified resulting in the linewidth increasing, which is observed in the RL spectrum and has been observed in the ASE spectrum \cite{howel04.01}.

\section{Conclusions}
Based on the observation of reversible photodegradation in DO11/PMMA and the desirability of robust organic-dye based random lasers for a variety of applications (such as speckle-free imaging \cite{Redding12.01}, tunable light sources \cite{Cao03.01} and optical physically unclonable functions \cite{Anderson14.04,Anderson14.05}), we investigate the emission properties of DO11+ZrO$_2$/PMMA under nanosecond optical pumping.  We find that for dye concentrations of less than 1.0 wt\%, DO11+ZrO$_2$/PMMA lases in the IFRL regime; while for a dye concentration of 1.0 wt\%, no lasing is observed for pump intensities up to the ablation threshold ($I\approx 754$ MW/cm$^2$).  The lasing threshold is found to increase with concentration, with the 0.1 wt\% sample having a threshold intensity of $75.8 \pm 9.4$ MW/cm$^2$ and the 0.5 wt\% sample having a threshold intensity of $121.1 \pm 2.1$ MW/cm$^2$. Both concentrations are found to have lasing wavelengths near 645 nm with a linewidth of approximately 10 nm. This lasing wavelength region is attractive for use with hydrocarbon based polymers as these polymers have an absorption minimum near 650 nm \cite{howel02.01}.

Along with considering the random lasing properties of DO11+ZrO$_2$/PMMA, we also measure the materials photodegradation and recovery.  We find that DO11+ZrO$_2$/PMMA photodegrades reversibly, with both the RL spectra before and after a photodegradation and recovery cycle being identical.  During photodegradation the lasing peak is found to blueshift, widen, and decrease in intensity, while during recovery the opposite process is found to occur with the lasing peak returning to its initial intensity, location, and linewidth.  This suggests that the observed degradation is truly reversible, which contrasts to measurements of R6G+NP/PU random lasers where the linewidth and peak wavelength are changed after decay and recovery \cite{Anderson15.01,Anderson15.03}.

While DO11+ZrO$_2$/PMMA is found to reversibly photodegrade like DO11/PMMA, the introduction of NPs into the dye-doped matrix is found to affect the decay and recovery rates, with DO11+ZrO$_2$/PMMA found to display increased photostability and to recover more quickly than similarly dye-doped DO11/PMMA.  These changes are explicable within the CCDM  \cite{raminithesis,andersonthesis,Ramini13.01, Anderson14.02}, with the NPs resulting in the free energy advantage increasing to an estimated value of 0.41 eV, but having little to no effect on the unitary domain decay and recovery rates, which are found to be in agreement with previous measurements of DO11/PMMA without NPs \cite{Ramini13.01}. While we have provided estimates of the CCDM model parameters for DO11+ZrO$_2$/PMMA, a more thorough study is required to determine the precise values.  Therefore we are planning measurements of DO11+ZrO$_2$/PMMA's decay and recovery for different temperatures, applied electric fields, and dye concentrations, which allow for the calculation of all CCDM parameters \cite{Ramini12.01,Ramini13.01,raminithesis,Anderson14.02}.

Finally, the observation of fully reversible photodegradation in DO11+ZrO$_2$/PMMA has promising prospects in the development of robust photodegradation resistant random lasers for real-world applications. We forsee using random lasers based on DO11+ZrO$_2$/PMMA in low duty-cycle applications such that the effects of photodegradation are mitigated by the self-healing mechanism of DO11+ZrO$_2$/PMMA, thus allowing prolonged use of the material. While we use a very low duty cycle during our recovery measurements -- to minimize photodegration and maximize self-healing -- we hypothesize that the material can function without significant photodegradation at higher duty cycles, dependent on the pump intensity.  Further work is required to determine the actual break even duty cycle at which photodegradation and self-healing are balanced.

\section{Acknowledgements}
This work was supported by the Defense Threat Reduction Agency, Award \# HDTRA1-13-1-0050 to Washington State University.

\bibliographystyle{osajnl} 
\bibliography{PrimaryDatabase,ASLbib}

\end{document}